\documentclass[aps,prb,twocolumn,showpacs,superscriptaddress,longbibliography,amsmath,amssymb,floatfix]{revtex4-2}
\usepackage[pdftex]{graphicx}
\usepackage[english]{babel}
\usepackage[usenames]{color}
\usepackage{colortbl}
\usepackage{dcolumn}
\usepackage{bm}
\usepackage{ulem}
\begin{document}
\title{Dynamics of the multiferroic LiCuVO$_4$ influenced by electric field}
\author{S. K. Gotovko}
\affiliation{P. L. Kapitza Institute for Physical Problems, Russian Academy of Sciences, 119334 Moscow, Russia}
\affiliation{National Research University Higher School of Economics, 101000 Moscow, Russia}
\author{V. I. Marchenko}
\affiliation{P. L. Kapitza Institute for Physical Problems, Russian Academy of Sciences,  119334 Moscow, Russia}
\author{A. Prokofiev}
\affiliation{Institute of Solid State Physics, Vienna University of Technology, 1040 Vienna, Austria}
\author{L. E. Svistov}
\email{svistov@kapitza.ras.ru}
\affiliation{P. L. Kapitza Institute for Physical Problems, Russian Academy of Sciences,  119334 Moscow, Russia}
\date{\today}
\begin{abstract}
We present the electron spin resonance study of the influence of an electric field on the low-field multiferroic magnetic state in LiCuVO$_4$. The shift of the magnetic resonance spectra in the electric field has been observed experimentally. Symmetry analysis has been conducted in order to describe the static properties of the magnetic system. The low-frequency dynamics of LiCuVO$_4$ in magnetic and electric fields was considered in the framework of hydrodynamic approach. It was shown that the application of the external electric field leads to the change of the configuration of the magnetic system before and after spin-flop reorientation. Satisfactory agreement was obtained between the results of experimental studies and theoretical consideration.
\end{abstract}

\maketitle

\section{Introduction}
There is a wide range of planar spiral magnets exhibiting multiferroic properties. Remarkable representatives of such materials are frustrated spin-chain compounds~\cite{park_2007,yasui_2011,yasui_2008,mourigal_2011,sergienko_2006,katsura_2005}.
A sufficiently low symmetry of spin ordering in these magnets leads to the appearance of a spontaneous electric polarization of the spin-orbit nature.
One of such magnets is LiCuVO$_4$ which is the object of the current research.

LiCuVO$_4$ is a member of the family of the frustrated spin-1/2 chain systems with competing nearest ferromagnetic and next-nearest neighbor antiferromagnetic exchange interactions \cite{enderle_2005}. Elastic neutron scattering experiments \cite{gibson_2004} revealed that an incommensurate cycloidal structure is established below ${T_N=2.3}$~K at magnetic fields
${\mu_0 H\lesssim 6.5}$~T. Due to the anisotropic magnetic susceptibility, orientation of the spin plane can be controlled by value and direction of external magnetic field. Electric polarization measurements, presented in Refs.~\cite{yasui_2008,schrettle_2008}, show that within the magnetoordering LiCuVO$_4$ in cycloidal phase acquires a spontaneous electric polarization $\boldsymbol{P},$ value and direction of which can be influenced by the application of magnetic field.

In this paper we discuss the results of a study of the low-frequency dynamics in LiCuVO$_4$ in the presence of an electric field obtained with the use of ESR technique.

Symmetry analysis has been conducted in order to describe the static properties of LiCuVO$_4$. Dynamic properties of LiCuVO$_4$ in the presence of an electric field have been given within the framework of phenomenological description~\cite{andreev_1980}.  

\section{Crystal and magnetic structure}

The crystal lattice of LiCuVO$_4$ belongs to the orthorhombic space group
$Imma.$ One unit cell contains four Cu$^{2+}$ ions (${a=5.662}$~\AA,
${b=5.809}$~\AA, ${c=8.758}$~\AA)~\cite{lafontaine_1989}. Copper ions form
chains running along the $b$-axis. Cu$^{2+}$ ions are surrounded by
edge-sharing octahedra of O$^{2-}$ ions. Such configuration of the oxygen surrounding results in the interesting case when exchange interactions between nearest and next nearest in-chain neighbors can be of the same order of value.

Results of elastic neutron scattering experiments~\cite{gibson_2004} in LiCuVO$_4$ show that an incommensurate spiral planar spin structure with the wave vector ${\boldsymbol{k}=(0,0.532,0)}$ is established below ${T_N=2.3}$~K. Spin plane of this structure lies in $ab$-plane of the crystal. Such structure, according to Ref.~\cite{enderle_2005}, is caused by the frustration of in-chain exchange interactions: ${J_N\approx-19}$~K and ${J_{NN}\approx45}$~K, and the exchange interaction between chains is one order weaker, that defines quasi-one dimensionality of the magnetic system. Note that the hierarchy of the exchange interactions in LiCuVO$_4$ up to date is an object of discussions~\cite{nishimoto_2012}.

Depending on direction and strength of the applied magnetic field, LiCuVO$_4$
demonstrates a diversity of magnetic phases. The application of the magnetic
field within the plane of cycloid induces spin-flop reorientation at
${\mu_0 H_{sf}\approx2.5}$~T~\cite{buttgen_2007}.
At a magnetic field ${\mu_0 H\gtrsim 6.5}$~T a number of exotic magnetic phases expected for quasi-one dimensional $J_N-J_{NN}$ magnets were observed~\cite{buttgen_2007,buttgen_2014,mourigal_2012,svistov_2011}.
 
In the experiments discussed in this paper, we studied the LiCuVO$_4$ in low field range where  planar cycloidal magnetic structure with the spontaneous electric polarization $\boldsymbol{P}$ occurs. Polarization appears at the Neel temperature and strongly depends on the direction and value of the applied field. 
At ${T=1.3}$~K and ${\mu_0 H=0}$ the value of electric polarization $P$ is approximately equal to~${35}$~$\mu$C/m$^2$~\cite{yasui_2008,schrettle_2008}. 

According to Ref.~\cite{andreev_1980}, three branches of low-frequency spin wave spectrum
are expected for a planar cycloidal structure. The branch that corresponds to the rotations of the spin plane around vector normal to it is gapless. Two other branches with gaps $\nu_{01}$ and $\nu_{02}$ correspond to the spin plane oscillations around $\boldsymbol{a}$ and $\boldsymbol{b}$ axes. In case of LiCuVO$_4$, the difference between the gaps was not detected, experimentally observed values are ${\nu_{01}\simeq\nu_{02}=\nu_0=30}$~GHz~\cite{buttgen_2007,prozorova_2016}.
In the experiments described below the influence of electric field on the antiferromagnetic resonance in LiCuVO$_4$ was studied. 

\section{Experiment}

\subsection{Technical details and methods}
Single crystals of LiCuVO$_4$ grown by the technique described in Refs.~\cite{prokofiev_2004,prokofiev_2005} have the shape of thin plates with developed $ab$-planes. The samples were of $1-3$~mm in $ab$-plane and $0.2-0.3$~mm in $\boldsymbol{c}$-direction. Some of the samples were from the same batches as ones used previously in NMR and electric polarization measurements by authors of~Refs.~\cite{schrettle_2008,ruff_2019}.

Electron spin resonance was measured with the use of a multiple-mode rectangular resonator of the transmission type in magnetic fields up to 7~T.
The temperature of the resonator with the sample was regulated from 1.3 to 10~K.

\begin{figure}[t]
	\includegraphics[width=0.8\columnwidth,angle=0,clip]{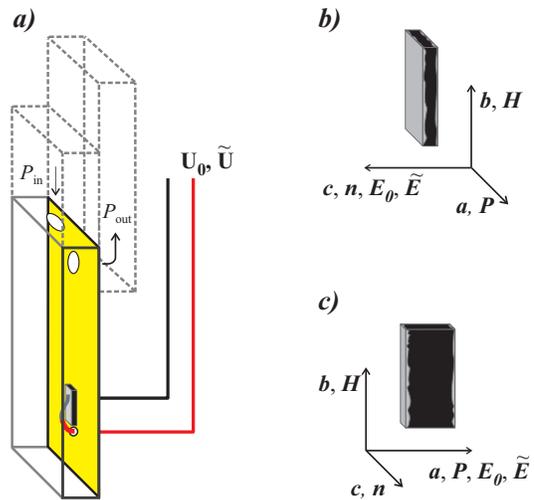}
	\caption{(Color online.) a) Scheme of the experimental cell: rectangular resonator with holes of coupling with the UHF-transmission line, the sample with a silver paste electrode. b, c) Schema of the samples and mutual orientations of the applied electric and magnetic fields $\boldsymbol{E}$ and $\boldsymbol{H},$ the electric polarization $\boldsymbol{P}$ from Refs.~\cite{schrettle_2008,yasui_2008} and crystallographic axes $\boldsymbol{a}$, $\boldsymbol{b}$, $\boldsymbol{c}$ for the cases $\boldsymbol{E}\parallel\boldsymbol{c}$ and $\boldsymbol{E}\parallel\boldsymbol{a}$, respectively.} \label{cell}
\end{figure}

 The crystal of LiCuVO$_4$ with electrodes made of silver paste was glued to the wall of the resonator as it is schematically shown in Fig.~\ref{cell}a). Such construction allows to apply the electric voltage up to 300~V between the electrodes before electrical breakdown. Presence of the electrodes and current conductors leads to essential reduction of the quality factor of the resonator and shields the sample from  the microwave field. Nevertheless, it was possible to observe absorption lines which reach $2-5$\% absorption of microwave power transmitted through the resonator.
  The value of the electric field in the sample can be roughly estimated as ${E\approx{U/d}}$ where $d$ is the distance between electrodes. Due to the plate form of the sample, the electric field in the main part of the sample for the $\boldsymbol{c}$-planes electrodes is uniform, whereas the electric field created by the electrodes applied to the $\boldsymbol{a}$-planes is expected to be essentially nonuniform. In the following description of the experiments, we give the value of $E$ assuming it being equal to $U/d$ as a rough estimation.
The mutual orientation of the crystallographic axes and the applied magnetic and electric fields for two (interesting for us) directions of electric field are shown in Fig.~\ref{cell}b),~c).

The electric field applied to the sample in our experiments was not strong enough to observe the shift of the resonance field ${{\Delta}H_{R}}$ in direct measurements of antiferromagnetic resonance.
For this reason the modulation method was used to study the influence of the electric field on the resonance curve. The application of an alternating electric field $\tilde{E}$ leads to the oscillations of $H_{R}$ and, as the result, to the oscillations of the transmitted through the resonator UHF-power. The oscillating
part of the transmitted power was measured by phase detection technique. Such a technique was used previously in Refs.~\cite{smirnov_1994,maisuradze_2012,gotovko_2018}.
Modulation frequency of $\tilde{E}$ was in the range of 200-600 Hz and the experimental results did not depend on modulation frequency.

In the absence of an external electric field, two energetically degenerate magnetic states with opposite directions of the electric polarization $\boldsymbol{P}$ are present in the sample.
Cooling of the sample from the paramagnetic state at application of sufficiently strong static electric field $\boldsymbol{E_0}\parallel\boldsymbol{P}$ removes this degeneracy. According to Refs.~\cite{schrettle_2008,yasui_2008}, the value of field $E_0=100$ kV/m is sufficient to prepare a single-domain sample of LiCuVO$_4$.

The strongest influence of the electric field on the magnetic resonance is expected in the vicinity of the spin wave spectra gap ${\nu_0=30}$~GHz.
The experiments discussed further were performed on the modes of the resonator in the range from 18 to 45~GHz.

\subsection{ESR Results}

\begin{figure}[t]
	\includegraphics[width=0.9\columnwidth,angle=0,clip]{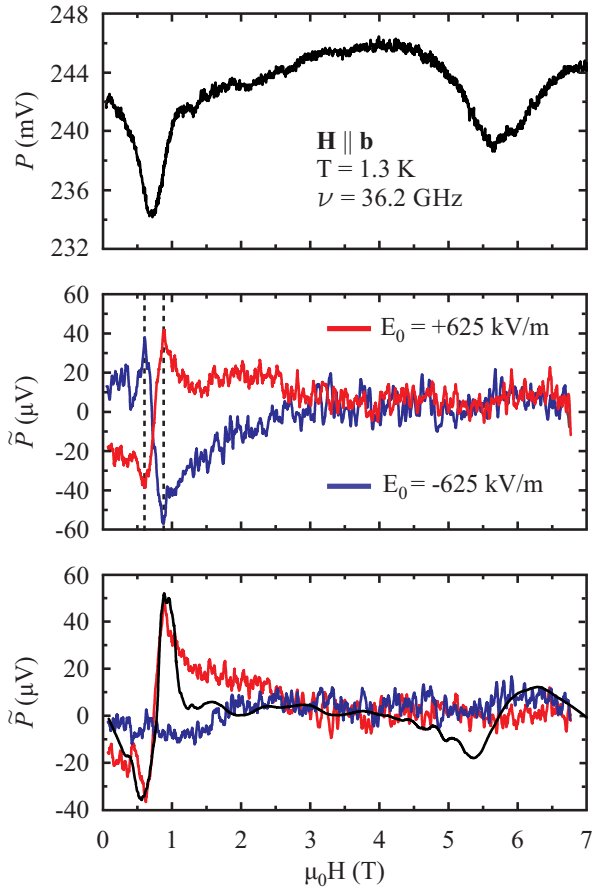}
	\caption{(Color online.) Upper panel: $\cal{P}$$(H)$-scan. Middle panel: $\tilde{\cal{P}}$$(H)$-scans measured at the permanent electric fields ${E_0}=+625$~kV/m and ${E_0}=-625$~kV/m (red and blue lines, respectively), amplitude of $\tilde{E}$ is equal to $375$~kV/m. Bottom panel: algebraic half-difference (red line) and half-sum (blue line) of $\tilde{\cal{P}}$$(H)$ presented in the middle panel. Black solid line represents field derivative of $\cal{P}$$(H)$ scaled with coefficient ${\alpha=18\cdot 10^{-4}}$~T.  ${\boldsymbol{H}\parallel\boldsymbol{b}}$, $\nu=36.2$~GHz, $T=1.3$~K.}
	\label{36_2}
\end{figure}

\begin{figure}[t]
	\includegraphics[width=0.9\columnwidth,angle=0,clip]{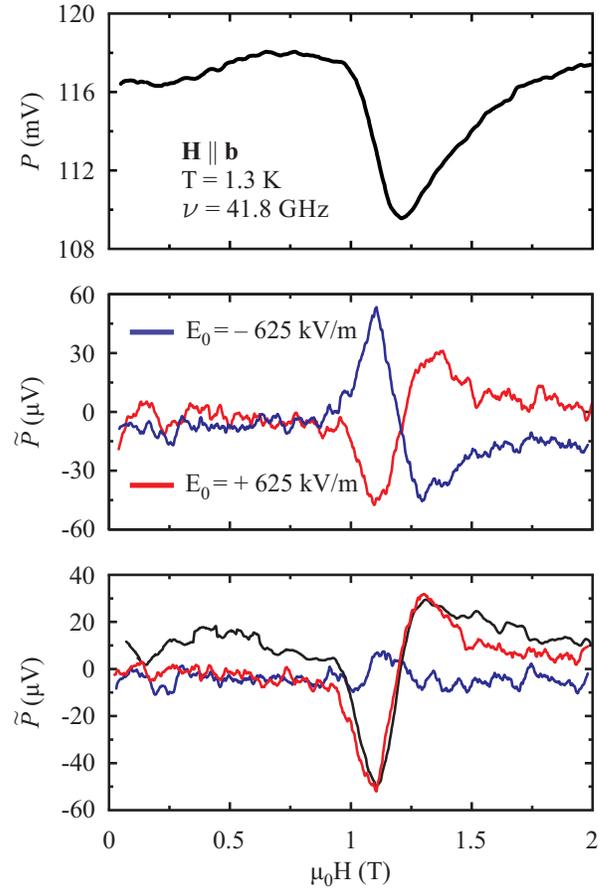}
	\caption{(Color online.) Upper panel: $\cal{P}$$(H)$-scan. Middle panel: $\tilde{\cal{P}}$$(H)$-scans measured at the permanent electric fields ${E_0}=+625$~kV/m and ${E_0}=-625$~kV/m (red and blue lines, respectively), amplitude of $\tilde{E}$ is equal to $375$~kV/m. Bottom panel: algebraic half-difference (red line) and half-sum (blue line) of $\tilde{\cal{P}}$$(H)$ presented in the middle panel. Black solid line represents field derivative of $\cal{P}$$(H)$ scaled with coefficient ${\alpha=10\cdot 10^{-4}}$~T. ${\boldsymbol{H}\parallel\boldsymbol{b}}$, $\nu=41.8$~GHz, $T=1.3$~K.}
	\label{41_8}
\end{figure}

Field scans of the transmitted through resonator UHF power $\cal{P}$ are shown in upper panels of Figs.~\ref{36_2},~\ref{41_8}. The measurements were performed at the temperature ${T=1.3}$~K, below temperature of the magnetic ordering (${T_N=2.3}$~K). Lines in Figs.~\ref{36_2},~\ref{41_8} were measured at resonance frequencies of multi-mode resonator ${\nu=36.2}$~GHz and  ${\nu=41.8}$~GHz, respectively. Static field $\boldsymbol{H}$ was applied along crystallographic $\boldsymbol{b}$-axis. Obtained resonance fields are in agreement with the results reported in Ref.~\cite{prozorova_2016}. Low field absorption lines are observed in the fields below spin-flop reorientation and correspond to the branch of magnetic resonance which rises quasi-linearly with field from $\nu_0$. The line measured at ${\nu=36.2}$~GHz demonstrates the second absorption line at the fields ${H>H_{sf}}$. The resonance fields at studied frequencies are marked in the frequency-field diagram in Fig.~\ref{spectra}. The middle panels of Figs.~\ref{36_2},~\ref{41_8} show the oscillating part of transmitted power $\tilde{\cal{P}}$ measured with the lock-in-amplifier. The measured signal $\tilde{\cal{P}}$ was in phase with alternating voltage $\tilde{U}$ used as reference for the amplifier. Before each recording, the sample was warmed up to the paramagnetic phase ($T=10$~K) and cooled down in a constant electric field $E_0$ created by the same electrodes as the alternating field $\tilde{E}$. During the recording, the electric field $E_0=625$~kV/m was not turned off. The field was used to keep the sample single-domain.
The records obtained at ${E_0 = + 625}$~kV/m and ${E_0 = - 625}$~kV/m are shown in Figs.~\ref{36_2},~\ref{41_8} with red and blue lines, respectively. The amplitude of the alternating electric field ${\tilde{E}}$ during the records was 375~kV/m. Both records $\tilde{\cal{P}}(H)$ in the range of low frequency absorption line reproduce the shape of the field derivative of $\pm{\cal{P}}(H)$. Positive (negative) sign corresponds to the case when the application of the positive electric field to the sample leads to the shift of the absorption line to lower (higher) fields. Such behavior of $\tilde{\cal{P}}$ corresponds to the shift of the absorption line to lower fields at the moments when the alternating electric field $\boldsymbol{\tilde{E}}(t)$ is co-directed with the permanent electric field $\boldsymbol{E_0}$. The lower panels of Figs.~\ref{36_2},~\ref{41_8} present the half-sum and half-difference of $\tilde{\cal{P}}(H)$ records obtained with positive and negative $\boldsymbol{E_0}$. For the case of single-domain sample the half-sum is equal to zero, whereas the half-difference of the signals can be fitted by a field derivative of ${\cal{P}}(H)$ shown in Figures with black lines.
Both $\cal{P}$ and $\tilde{\cal{P}}$ presented in Figures are measured in arbitrary but the same units, that allows determining the value of the shift of the resonance field for absorption lines. The shift of resonance line $\Delta{\mu_0 H}_{R}$ at application of amplitude value of alternating electric field ${{E}=375}$~kV/m is equal to the double scaling factor. At $E=375$~kV/m, $\Delta{\mu_0 H}_{R}$ is equal to $(36\pm5)\cdot10^{-4}$~T for  ${\nu=36.2}$~GHz, and $\Delta{\mu_0 H}_{R}$ is equal to $(20\pm5)\cdot10^{-4}$~T for ${\nu=41.8}$~GHz. In the range of the resonance absorption line observed at $H>H_{sf}$ we did not observe any response of transmitted power on the applied electric field.

\begin{figure}[t]
\includegraphics[width=0.9\columnwidth,angle=0,clip]{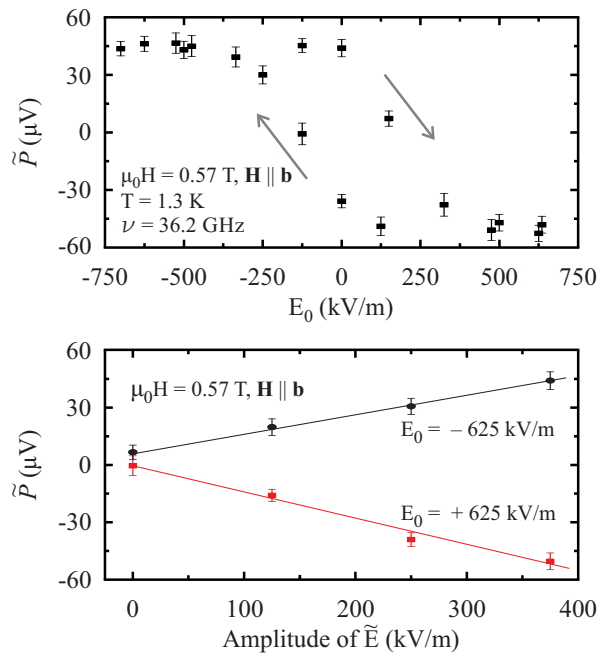}
\caption{(Color online.) Upper panel: $\tilde{\cal{P}}$$(E_0)$. Arrows indicate direction in which the value of $E_0$ was changed. Amplitude of $\tilde{E}$ is equal to $375$~kV/m. Bottom panel: dependence of $\tilde{\cal{P}}$ on amplitude of $\tilde{E}$ measured at $E_0=\pm625$~kV/m. ${\boldsymbol{H}\parallel\boldsymbol{b}}$, $\mu_0H=0.57$~T, $\nu=36.2$~GHz, and $T=1.3$~K.}
\label{36_2_points}
\end{figure}

Fig.~\ref{36_2_points} presents $\tilde{\cal{P}}$ at ${\nu=36.2}$~GHz measured at field ${\mu_0 H=0.57}$~T at the extremum of the response on alternating electric field. The amplitudes were measured with high integration time of the lock-in-amplifier. Upper panel shows dependence of $\tilde{\cal{P}}$ on the value of the external permanent electric field $E_0$.
Amplitude of the alternating electric field ${\tilde{E}}$ was 375~kV/m; value
of $E_0$ was gradually changed as it is shown in Figure with gray arrows.
This figure demonstrates that $\tilde{\cal{P}}$ saturates at
${E_0>400}$~kV/m, therefore the sample is electrically single-domain at higher electric fields. Bottom panel demonstrates linearity of
$\tilde{\cal{P}}$ on amplitude of the external alternating electric field
$\tilde{E}$ at ${E_0=\pm625}$~kV/m. The observed response of transmitted power on alternating electric field did not change at the switch of the direction of the static magnetic field $\boldsymbol{H}$. These two observations are important for excluding the effect of  magnetoelectric coupling discussed in Refs.~\cite{smirnov_1994,vitebskii_1991}.

\begin{figure}[t!]
	\includegraphics[width=0.9\columnwidth]{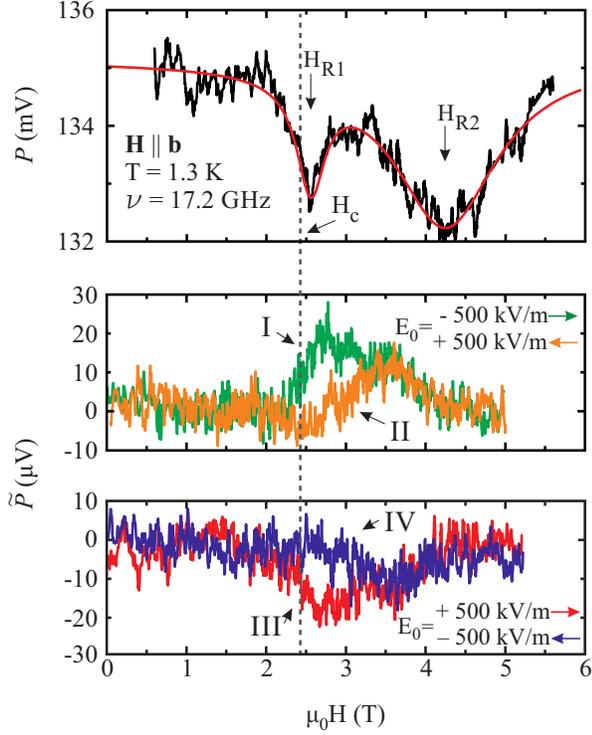}
	\caption{(Color online.) Upper panel: ${\cal{P}}(H)$-scan (black line) and its interpolation by two Lorentzian line shape functions. Bottom panels: $\tilde{\cal{P}}$ at the frequency of the applied alternating electric field $\tilde{E}$ measured with the phase detection technique (red, orange, green, and blue lines). Directions of scanning and $E_0$ at which $\tilde{\cal{P}}$ were measured are indicated in Figure by arrows with corresponding colors, order of arrows coincides with the measurement order. ${\boldsymbol{H}\parallel\boldsymbol{b}},$ ${\tilde{E}=250}$~kV/m. ${\nu=17.2}$~GHz, ${T=1.3}$~K.}
	\label{17_2}
\end{figure}

Fig.~\ref{17_2} presents field dependencies of $\cal{P}$ and $\tilde{\cal{P}}$ measured at ${\nu=17.2}$~GHz (below $\nu_0$), ${\boldsymbol{H}\parallel\boldsymbol{b}}.$ At this frequency two absorption lines are observed which are indicated in upper panel by arrows. One absorption line corresponds to the decreasing branch of the $\nu-H$ diagram at $H_{sf}$ and the second line corresponds to the rising one (see Fig.~\ref{spectra}).

Field dependencies of $\tilde{\cal{P}}$ shown by lines I and III in the middle and upper panels of Fig.~\ref{17_2} were measured at the increase of magnetic field $\mu_0 H$ from zero to $5$~T at negative and positive signs of the permanent electric field $E_0$, respectively. The electric field was applied at zero magnetic field and was not switched during the scans. The observed signs of alternating responses correspond to the shift of high field absorption line $H_{R2}$ to the lower magnetic fields at the application of electric field along the polarization of the sample.
The lines II and IV were obtained at magnetic field decrease after switching of the sign of the electric field $E_0$ to opposite one at $\mu_0 H\approx5$~T. The lines II and IV coincide
with the lines I and III down to $\mu_0 H=3.5$~T. At lower fields ${\mu_0 H}\lesssim3.5$~T the lines II and IV tend to zero. This experiment shows that electric depolarization of the sample takes place at fields ${H}$ close to $H_{sf}$ only. This observation seems to be natural taking into account that for ${\boldsymbol{H}\parallel\boldsymbol{b}}$ the magnetic spin-flopped phase is not polar \cite{schrettle_2008}.

\begin{figure}[t]
	\includegraphics[width=0.9\columnwidth,angle=0,clip]{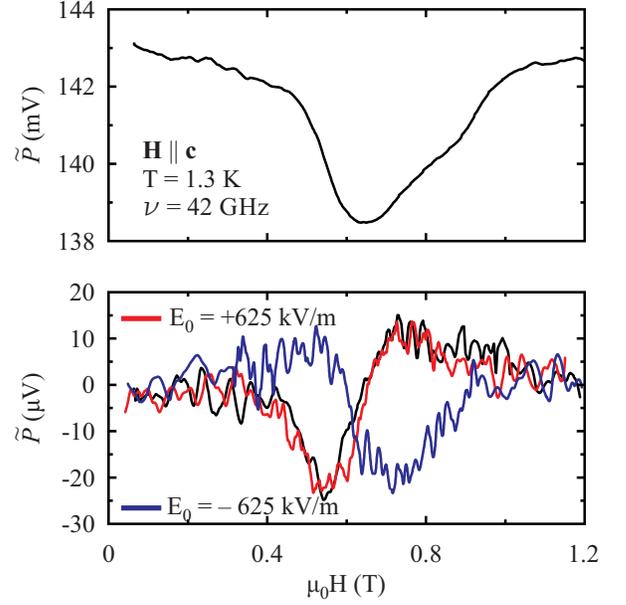}
	\caption{(Color online.) Upper panel: ${\cal{P}}(H)$-scan. Bottom panel: $\tilde{\cal{P}}$ at the frequency of $\tilde{E}$ measured with the phase detection technique (red and blue lines), the black solid line shows $d{\cal{P}}/dH$ scaled by the coefficient ${\alpha=7\cdot 10^{-4}}$~T. Amplitude of $\tilde{E}$ is equal to $375$~kV/m. ${\boldsymbol{H}\parallel\boldsymbol{c}},$ ${\nu=42}$~GHz, ${T=1.3}$~K.} 	 \label{42c}
\end{figure}

Fig.~\ref{42c} shows field dependencies of  the $\cal{P}$ and $\tilde{\cal{P}}$ at $\boldsymbol{H}\parallel \boldsymbol{c}$ and $\boldsymbol{E} \parallel \boldsymbol{a}$.
 Designations adopted in Fig.~\ref{42c} coincide with the designations in Figs.~\ref{36_2},~\ref{41_8}. For these orientations of the fields the response $\tilde{\cal{P}}$ was observed only within low field phase before spin-flop reorientation. This response corresponds to the shift of resonance field $\mu_0 H_R$ at application of electric field ${{E}=375}$~kV/m by the value $\Delta{\mu_0 H}_{R}=(14\pm5)\cdot10^{-4}$~T for ${\nu=42}$~GHz.

Experiments carried out for field orientation  $\boldsymbol{B}\parallel \boldsymbol{a,b}$ and  $\boldsymbol{E}\parallel \boldsymbol{c}$ (see  Fig.~\ref{cell}b) did not detect any shift of resonance field at applied electric field.

\section{Theory}

The wave vector of the magnetic structure of LiCuVO$_4$ is ${(0,0.532,0)}$
which is close to the wave vector of a commensurate structure ${(0,0.5,0)}.$
Following Dzyaloshinskii~\cite{dzyalosh_1957}, at first we consider the
magnetic transition to the commensurate state. Incommensurability in case of
LiCuVO$_4$ is stipulated by presence of a small Lifshitz invariant.

\begin{figure}[t]
	\includegraphics[width=0.8\columnwidth,angle=0,clip]{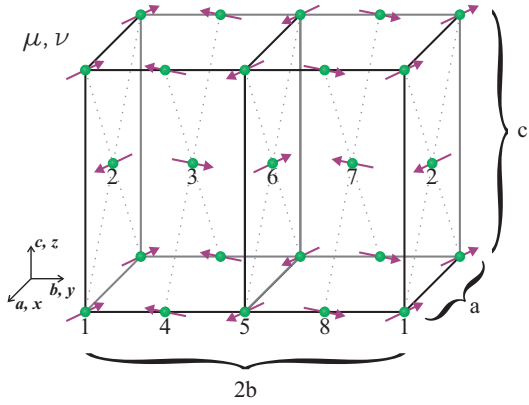}
	\caption{(Color online.) One magnetic cell of LiCuVO$_4.$ Only Cu$^{2+}$ ions
		are shown by green circles. Eight magnetic sublattices are denoted by numbers. Magnetic moments
		are presented for representation~(\ref{munu}) by arrows.} 	
	\label{magncell}
\end{figure}

We can point out eight magnetic sublattices. According to
Ref.~\cite{lafontaine_1989}, spins of Cu$^{2+}$ magnetic ions of these
sublattices within one magnetic cell have the following positions in units of
lattice parameters (See~Fig.~\ref{magncell}): $\boldsymbol{s}_1(0,0,0.25);$
$\boldsymbol{s}_2(0.5,0,-0.25);$ $\boldsymbol{s}_3(0.5,0.5,-0.25);$
$\boldsymbol{s}_4(0,0.5,0.25);$ $\boldsymbol{s}_5(0,1,0.25);$ $\boldsymbol{s}_6(0.5,1,-0.25);$
$\boldsymbol{s}_7(0.5,1.5,-0.25);$ $\boldsymbol{s}_8(0,1.5,0.25).$
Here we introduce axes $\boldsymbol{x}$, $\boldsymbol{y}$, and $\boldsymbol{z}$ which coincide with crystallographic axes $\boldsymbol{a}$, $\boldsymbol{b}$, and $\boldsymbol{c}$, respectively. 

The symmetry space group $Imma$ to which crystals of LiCuVO$_4$ belong consists
of the following transformations: translations
\begin{equation}\begin{split}
\boldsymbol{a}_1(x,y,z)=(x+0.5,y+0.5,z+0.5);\\
\boldsymbol{a}_2(x,y,z)=(x+0.5,y-0.5,z+0.5);\\
\boldsymbol{a}_3(x,y,z)=(x+0.5,y+0.5,z-0.5);
\end{split}\end{equation}
three rotations, and inversion
\begin{equation}\begin{split}
u_1(x,y,z)=(x,-y,-z+0.5);\\
u_2(x,y,z)=(-x+0.5,y,-z);\\
C_2(x,y,z)=(-x,-y+0.5,z);\\
I(x,y,z)=(-x,-y,-z+0.5).
\end{split}\end{equation}

These crystallographic transformations perform the following transpositions of
sublattices:
\begin{equation}\label{sublattices}\begin{split}
u_1:\, &\boldsymbol{s}_1\rightarrow\boldsymbol{s}_1, \,\boldsymbol{s}_2\rightarrow\boldsymbol{s}_2,
\boldsymbol{s}_3\rightarrow\boldsymbol{s}_7\rightarrow\boldsymbol{s}_3,\\
&\boldsymbol{s}_4\rightarrow\boldsymbol{s}_8\rightarrow\boldsymbol{s}_4,
\,\boldsymbol{s}_5\rightarrow\boldsymbol{s}_5; \boldsymbol{s}_6\rightarrow\boldsymbol{s}_6;\\ u_2:\, &
\boldsymbol{s}_1\rightarrow\boldsymbol{s}_2\rightarrow\boldsymbol{s}_1,
\,\boldsymbol{s}_3\rightarrow\boldsymbol{s}_4\rightarrow\boldsymbol{s}_3,\\
&\boldsymbol{s}_5\rightarrow\boldsymbol{s}_6\rightarrow\boldsymbol{s}_5,
\,\boldsymbol{s}_7\rightarrow\boldsymbol{s}_8\rightarrow\boldsymbol{s}_7;\\ C_2:\,
&\boldsymbol{s}_1\rightarrow\boldsymbol{s}_4\rightarrow\boldsymbol{s}_1,
\,\boldsymbol{s}_2\rightarrow\boldsymbol{s}_3\rightarrow\boldsymbol{s}_2,\\
&\boldsymbol{s}_5\rightarrow\boldsymbol{s}_8\rightarrow\boldsymbol{s}_5,
\,\boldsymbol{s}_6\rightarrow\boldsymbol{s}_7\rightarrow\boldsymbol{s}_6;\\ I:\,
&\boldsymbol{s}_1\rightarrow\boldsymbol{s}_1, \,\boldsymbol{s}_2\rightarrow\boldsymbol{s}_2,
\,\boldsymbol{s}_3\rightarrow\boldsymbol{s}_7\rightarrow\boldsymbol{s}_3,\\
&\boldsymbol{s}_4\rightarrow\boldsymbol{s}_8\rightarrow\boldsymbol{s}_4,
\,\boldsymbol{s}_5\rightarrow\boldsymbol{s}_5, \,\boldsymbol{s}_6\rightarrow\boldsymbol{s}_6;\\
\boldsymbol{a}_1:\, &\boldsymbol{s}_1\rightarrow\boldsymbol{s}_3\rightarrow\boldsymbol{s}_5
\rightarrow\boldsymbol{s}_7\rightarrow\boldsymbol{s}_1,\\
&\boldsymbol{s}_2\rightarrow\boldsymbol{s}_4\rightarrow\boldsymbol{s}_6
\rightarrow\boldsymbol{s}_8\rightarrow\boldsymbol{s}_2.
\end{split}\end{equation}

Let us introduce the following linear combinations:
\begin{equation}\begin{split}
\boldsymbol{s}&=\boldsymbol{s}_1+\boldsymbol{s}_2+\boldsymbol{s}_3+\boldsymbol{s}_4
+\boldsymbol{s}_5+\boldsymbol{s}_6+\boldsymbol{s}_7+\boldsymbol{s}_8,\\
\boldsymbol{\ell}_1&=\boldsymbol{s}_1+\boldsymbol{s}_2-\boldsymbol{s}_3-\boldsymbol{s}_4+\boldsymbol{s}_5+\boldsymbol{s}_6
-\boldsymbol{s}_7-\boldsymbol{s}_8,\\
\boldsymbol{\ell}_2&=\boldsymbol{s}_1-\boldsymbol{s}_2+\boldsymbol{s}_3-\boldsymbol{s}_4
+\boldsymbol{s}_5-\boldsymbol{s}_6+\boldsymbol{s}_7-\boldsymbol{s}_8,\\
\boldsymbol{\ell}_3&=\boldsymbol{s}_1-\boldsymbol{s}_2-\boldsymbol{s}_3+\boldsymbol{s}_4
+\boldsymbol{s}_5-\boldsymbol{s}_6-\boldsymbol{s}_7+\boldsymbol{s}_8,\\
\boldsymbol{\ell}_4&=\boldsymbol{s}_1+\boldsymbol{s}_2-\boldsymbol{s}_5-\boldsymbol{s}_6,\\
\boldsymbol{\ell}_5&=\boldsymbol{s}_3+\boldsymbol{s}_4-\boldsymbol{s}_7-\boldsymbol{s}_8,\\
\boldsymbol{\ell}_6&=\boldsymbol{s}_1-\boldsymbol{s}_2-\boldsymbol{s}_5+\boldsymbol{s}_6,\\
\boldsymbol{\ell}_7&=\boldsymbol{s}_3-\boldsymbol{s}_4-\boldsymbol{s}_7+\boldsymbol{s}_8.
\end{split}\end{equation}

According to the transformation rules (Eq.~(\ref{sublattices})), the ferromagnetic
vector $\boldsymbol{s},$ and antiferromagnetic vectors
$\boldsymbol{\ell}_1,\boldsymbol{\ell}_2,\boldsymbol{\ell}_3$ are transformed
by one-dimensional representations. Couples of antiferromagnetic vectors
$(\boldsymbol{\ell}_4,\boldsymbol{\ell}_5)$ and
$(\boldsymbol{\ell}_6,\boldsymbol{\ell}_7)$ are transformed by two-dimensional
representations. Both couples correspond to the wave vector
${\boldsymbol{k}=(0,0.5,0)}.$ According to the results of experimental study of
magnetic structure~(Ref.~\cite{gibson_2004}) only the couple of spin vectors
$(\boldsymbol{\ell}_6,\boldsymbol{\ell}_7)$ should be considered as an active
representation in Dzyaloshinskii-Landau theory of antiferromagnetic second
order phase transition in LiCuVO$_4.$

To shorten the equations, we change the notation ${(\boldsymbol{\ell}_6,\boldsymbol{\ell}_7)
\rightarrow(\boldsymbol{\mu},\boldsymbol{\nu})}.$ These vectors are transformed as \mbox{follows}:
\begin{equation}\label{munu}\begin{split}
u_1:& \,\boldsymbol{\mu}\rightarrow\boldsymbol{\mu},
\boldsymbol{\nu}\rightarrow-\boldsymbol{\nu};\\ u_2:&
\,\boldsymbol{\mu}\rightarrow-\boldsymbol{\mu},
\boldsymbol{\nu}\rightarrow-\boldsymbol{\nu};\\ C_2:& \,
\boldsymbol{\mu}\rightarrow-\boldsymbol{\nu}\rightarrow\boldsymbol{\mu};\\ I:&
\,\boldsymbol{\mu}\rightarrow\boldsymbol{\mu},
\boldsymbol{\nu}\rightarrow-\boldsymbol{\nu};\\ \boldsymbol{a}_1:& \,
\boldsymbol{\mu}\rightarrow\boldsymbol{\nu}\rightarrow-\boldsymbol{\mu}.
\end{split}\end{equation}

Let us investigate the exchange invariants first. Invariants of 4th order
defining the structure of phases in case of the representation (\ref{munu}) are as follows:
\begin{equation}
B_1(\boldsymbol{\mu}\boldsymbol{\nu})^2+B_2(\boldsymbol{\mu}^2-\boldsymbol{\nu}^2)^2.
\end{equation}
Detected in Ref.~\cite{gibson_2004} the plane spin structure
${\boldsymbol{\mu}\bot\boldsymbol{\nu}},$ ${\boldsymbol{\mu}^2=\boldsymbol{\nu}^2}$ occurs at condition ${B_1>0},$ ${B_2>0}.$ Further we consider $\boldsymbol{\mu},$ $\boldsymbol{\nu}$ as unit vectors.

Exchange Lifshitz invariant is:
\begin{eqnarray}
C_L(\boldsymbol{\mu}\partial_y\boldsymbol{\nu}-\boldsymbol{\nu}\partial_y\boldsymbol{\mu}).
\end{eqnarray}
The proximity of the magnetic structure to a commensurate one is possible when $C_L$ is small.

Main relativistic invariants are ${\mu_x^2+\nu_x^2}$ and ${\mu_y^2+\nu_y^2}.$ Let us introduce unit vector in spin space ${\boldsymbol{n}=\boldsymbol{\mu}\times\boldsymbol{\nu}}$. Due to the identity
${n_in_k+\mu_i\mu_k+\nu_i\nu_k=\delta_{ik}},$  the anisotropy energy can be written in the form
\begin{equation}\label{Ua}
U=\frac{\beta_1}{2}n_x^2+\frac{\beta_2}{2}n_y^2.
\end{equation}

Note that this anisotropy fixes the orientation of vector $\boldsymbol{n},$ but orientation of the couple $(\boldsymbol{\mu},~\boldsymbol{\nu})$ inside spin plane remains free. So there is no competition with the Lifshitz invariant in a considered approximation. Observed equilibrium orientation of the spin structure corresponds to the case ${\beta_1,\beta_2>0}.$

In the applied electric field the following relativistic invariants appear:
\[(\mu_x\nu_y-\mu_y\nu_x)E_x~\equiv~n_zE_x,\,(\mu_y\nu_z-\mu_z\nu_y)E_z~\equiv~n_xE_z.\] Accordingly, it is necessary to add to the anisotropy energy (\ref{Ua}) the following terms:
\begin{equation}\label{UaE}
U_{aE}=-\lambda_1n_zE_x-\lambda_2n_xE_z,  	
\end{equation}
that means the occurrence of two components of the spontaneous electric polarization in the antiferromagnetic phase:
\begin{equation}
P_x=\lambda_1n_z,\, P_z=\lambda_2n_x.
\end{equation}

Microscopic consideration of possible mechanisms of spontaneous electric polarization in crystals of LiCuVO$_4$ is presented in Refs.~\cite{katsura_2005,eremin_2019}.

Using the equations of low-frequency spin dynamics~\cite{andreev_1980}, we obtain three branches of spin waves in the planar antiferromagnet LiCuVO$_4,$ corresponding to oscillations of the unit vector $\boldsymbol{n}$ and rotation of the spin structure around $\boldsymbol{n}.$ The last degree of freedom is gapless. Two frequencies of dynamics of the vector $\boldsymbol{n}$ at zero wave vector in case of ${\boldsymbol{H}\parallel{\boldsymbol{y}}}$ and ${\boldsymbol{E}\parallel{\boldsymbol{x}}}$ are defined by the biquadratic equation:
\begin{widetext}
\begin{equation}\label{freqeq}\begin{split}
\nu^4-\nu^2\left\{(3\cos^2\alpha-2)\nu^2_{10}+\nu^2_{20}+(1+\eta^2\sin^2\alpha)\gamma^2H^2+ 2\epsilon\cos\alpha\right\}+\\
+\left\{\nu^2_{10}\cos2\alpha+(\eta\sin^2\alpha+\cos^2\alpha)\gamma^2H^2+
\epsilon\cos\alpha\right\}\left\{\nu^2_{20}-\nu^2_{10}\sin^2\alpha+
\eta\sin^2\alpha\gamma^2H^2+\epsilon\cos\alpha\right\}=0.
\end{split}\end{equation}
\end{widetext}
Here ${\eta=(\chi_{\parallel}-\chi_{\perp})/\chi_{\perp}},$ $\alpha$ is the angle between
$\boldsymbol{n}$ and $z,$
\[\nu_{10}=\gamma\sqrt{\frac{\beta_1}{\chi_{\perp}}},
\, \nu_{20}=\gamma\sqrt{\frac{\beta_2}{\chi_{\perp}}}, \,
\epsilon=\gamma^2\frac{P_xE_x}{\chi_{\perp}}, \gamma=g\frac{\mu_B}{h},\]
where $\mu_B$ is the Bohr magneton, $h$ is Planck's constant, and ${g=2}$ is the $g$-factor of free electron. Note that in the framework of the hydrodynamic theory~(Ref.~\cite{andreev_1980}) corrections to $g$-factor should be described with the aid of special relativistic invariants in Lagrange procedure.

The value of $\alpha$ is defined by the following expressions:
\begin{equation}\label{alpha}\begin{split}
H<H^{*}&=\sqrt{H_{sf}^2+\frac{P_xE_x}{\chi_\parallel-\chi_\perp}}:\, \alpha=0,
\\ H>H^{*}&:\, \cos\alpha=\frac{P_xE_x}{\eta\chi_{\perp}(H^2-H_{sf}^2)},
\end{split}\end{equation}
here ${H_{sf}=\sqrt{\beta_1/\eta\chi_{\perp}}}.$

When ${E=0},$ ${H<H_{sf}},$ Eq.~(\ref{freqeq}) transforms into
\begin{equation}\label{freq0}
\nu_1=\sqrt{\nu_{10}^2+\gamma^2H^2}, \, \nu_2=\nu_{20}.
\end{equation}

According to Ref.~\cite{buttgen_2007}, ${\nu_1=\nu_2\equiv\nu_0}.$ 	

At ${\boldsymbol{H}\parallel{\boldsymbol{x,y}},~H>H_{sf}},$ ${E=0}$ the frequencies are defined by the following expression:
\begin{equation}\label{freqsf0sf}\begin{split}
\nu_{1,2}^2=\frac{(1+\eta^2)}{2}\gamma^2H^2-\frac{\nu_{0}^2}{2} \pm\\ \pm
\frac{1}{2} \sqrt{\nu_{0}^4 -2\nu_{0}^2(\eta-1)^2\gamma^2H^2
+(\eta^2-1)^2\gamma^4H^4}.
\end{split}\end{equation}

At ${\boldsymbol{H}\parallel{\boldsymbol{x,y}},~H>H_{sf}},$ ${\boldsymbol{E}\parallel\boldsymbol{x}}$ the expression for the frequencies can be calculated numerically with the use of Eq.~(\ref{freqeq}).

At ${\boldsymbol{H}\parallel{\boldsymbol{x}}}, {H>H_{sf}},~{\boldsymbol{E}\parallel{\boldsymbol{z}}}$ the frequencies are
\begin{equation}\label{freqsfEsfa}\begin{split}
		\nu_{1,2}^2=\frac{(1+\eta^2)}{2}\gamma^2H^2-\frac{\nu_{0}^2}{2}+\gamma^2\frac{E_zP_z}{\chi_\perp} \pm\\ \pm
		\frac{1}{2} \sqrt{\nu_{0}^4 -2\left( \nu_{0}^2-2\gamma^2\frac{E_zP_z}{\chi_\perp}\right) (\eta-1)^2\gamma^2H^2
			+(\eta^2-1)^2\gamma^4H^4}.
\end{split}\end{equation}

For ${\boldsymbol{H}{\parallel}\boldsymbol{z}},$ ${\boldsymbol{E}{\parallel}\boldsymbol{x}}$ the spectra is
\begin{equation}\label{freqc}\begin{split}
\nu_{1,2}^2=\nu_0^2+\gamma^2\frac{1+\eta^2}{2}H^2+\epsilon \, \pm\\
\pm(1-\eta)\gamma{H}\sqrt{\nu_0^2+ \frac{(1+\eta)^2}{4}\gamma^2H^2+\epsilon}.
\end{split}\end{equation}

\section{Discussion of experimental results}
In our experiments both permanent and alternating  electric fields ($E_0$ and $\tilde{E}$) were applied to the sample. Alternating response of the transmitted through resonator UHF power $\tilde{\cal{P}}$ on the electric field was studied. The dependence of $\tilde{\cal{P}}(E_0)$ on permanent electric field saturates at $E_0>400$~kV/m (see Fig.~\ref{36_2_points}). It means that at high electric field samples of LiCuVO$_4$ are electrically polarized, i.e., only one magnetic domain presents in the sample due to the interaction with electric field. Transmitted through the resonator power ${\cal{P}}(H)$ for the case of single-domain sample can be written as

\begin{equation}
{\cal{P}}(H,E_0+\tilde{E})={\cal{P}}(H,0)+
\frac{\partial{\cal{P}}}{\partial{H}}
\frac{\partial{H}_R}{\partial{E}}(E_0+\tilde{E}). \label{transpower}
\end{equation}

The low-field ESR frequency $\nu_1(H,E)$ at
${\boldsymbol{H}\parallel{\boldsymbol{y}},\,\boldsymbol{E}\parallel{\boldsymbol{x}}},$ ${H<H^{*}}$ is defined by the following expression that can be obtained from Eq.~(\ref{freqeq}):
\begin{equation}
 \nu_1^2=\nu_{10}^2+\gamma^2H^2+\epsilon.
 \label{lowfield}
\end{equation}
Here we remind that axes $\boldsymbol{x}$, $\boldsymbol{y}$, $\boldsymbol{z}$ are directed along crystallographic axes $\boldsymbol{a}$, $\boldsymbol{b}$, $\boldsymbol{c}$.
Assuming that $E_x$ is small, we obtain the shift of $H_{R}$ from Eq.~(\ref{freqc}) for ${\boldsymbol{H}\parallel{\boldsymbol{y}}}$ and ${\boldsymbol{H}\parallel{\boldsymbol{z}}}$ for rising branches:
\begin{equation}
\Delta{H}_{R}(\boldsymbol{H}\parallel{\boldsymbol{y}})=-\frac{P_x E_x}{2\chi_{\perp}H_{R}}, \label{deltaY}
\end{equation}
\begin{equation}\label{deltaZ}
\Delta{H}_{R}(\boldsymbol{H}\parallel{\boldsymbol{z}})=
-\frac{\gamma^2P_xE_x}{2\chi_{\perp}\eta\gamma_c^2H_{R}
+\chi_{\perp}\gamma_c(1-\eta)\nu}.
\end{equation}

\begin{figure}
	\includegraphics[width=0.9\columnwidth,angle=0,clip]{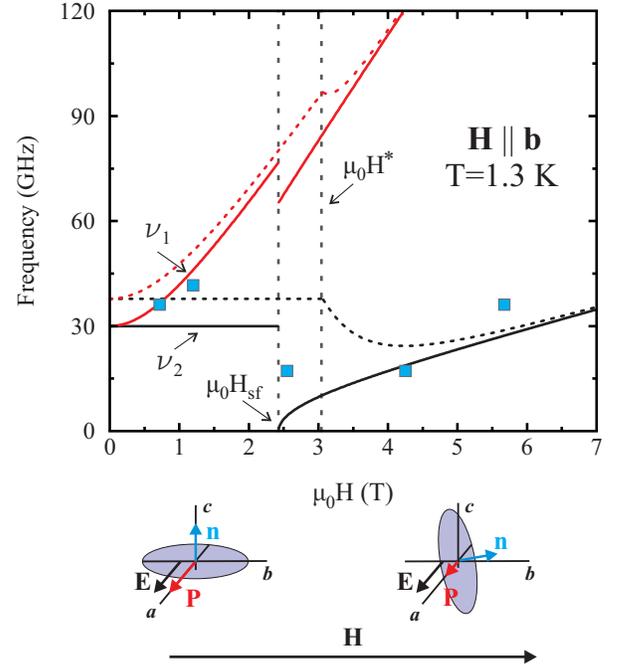}
	\caption{(Color online.) Frequency--resonance-field diagrams computed for magnetic field $\boldsymbol{H}\parallel\boldsymbol{b}$. Solid lines: $\nu_{1,2}(H_{R})$ at zero electric field, dashed lines: $\nu_{1,2}(H_{R})$ at ${E_0=50}$~MV/m. Blue solid squares show experimental values of $\nu(H_{R})$ obtained in experiments discussed in this paper.}
	\label{spectra}
\end{figure}

In Eqs.~(\ref{lowfield})-(\ref{deltaZ}), the positive sign of $P_xE_x$ corresponds to the case when the external electric field is co-directed with the electric polarization. From Eqs.~(\ref{transpower})-(\ref{deltaZ}) we obtain the expected value of oscillating part of the transmitted power $\tilde{\cal{P}}$ for rising branches:
\begin{equation}
\tilde{\cal{P}}(\boldsymbol{H}\parallel{\boldsymbol{y}})=\frac{\partial{\cal{P}}}{\partial{H}}
\frac{\partial{H}_R}{\partial{E}}\tilde{E}=
-\frac{\partial{\cal{P}}}{\partial{H}}\frac{P_x}{2\chi_{\perp}H_{R}}\tilde{E},
\label{ptry}
\end{equation}
\begin{equation}
\tilde{\cal{P}}(\boldsymbol{H}\parallel{\boldsymbol{z}})=
-\frac{\partial\cal{P}}{\partial{H}}\frac{\gamma^2P_x}
{2\chi_{\perp}\eta\gamma_c^2 H_{R}+\chi_{\perp}\gamma_c(1-\eta)\nu}\tilde{E}.
\label{ptrz}
\end{equation}

The frequency--resonance-field diagram computed for the model described by
Eq.~(\ref{freqeq}) is shown in Fig.~\ref{spectra}. Following parameters were
used: ${\chi_{\parallel}/\chi_{\perp}=1.18}$ and ${\nu_{01,02}=30}$~GHz
\cite{buttgen_2007, prozorova_2016}. These dependencies are shown with solid lines. The $\nu_2(B)$ dependence demonstrates an abrupt jump at the spin-flop field $H_{sf}$. 

The dependencies $\nu(H_{R})$ computed for ${E=50}$~MV/m, ${P_x=20}$~$\mu$C/m$^2$ are presented by dashed lines to illustrate the influence of the external electric field $E$ on the ESR spectra. Such value of $E$ could not be reached in our experiments and has been chosen for visual clarity (value of $E$ used in the experiments did not exceed  $750$~kV/m to avoid electrical breakdown). It can be seen that with the application of the electric field the shift of the ESR spectra is assumed, and for low-field part of the branch $\nu_1$ the effect is expected to be higher when the frequency of measurement $\nu$ is closer to the gap $\nu_{01}.$

As follows from the theoretical model, at the magnetic field higher than $H^{*}$ in absence of an electric field the magnetic structure undergoes spin-flop reorientation of the spin plane with $\boldsymbol{n}\parallel{\boldsymbol{c}}$ to $\boldsymbol{n}\parallel{\boldsymbol{b}}$. In the presence of an electric field the spin plane starts continuous rotation at fields $H>H^{*}$. The angle between $\boldsymbol{n}$ and $\boldsymbol{H}$ is defined by Eq.~(\ref{alpha}). The application of the electric field  is accompanied by contribution to the frequency gap of the magnetic resonance spectra and by the shift of $H^{*}$ to higher fields. Despite the fact that the influence of the electric field on $\nu_0$ and $H^{*}$ is similar to the change of the anisotropy constant, the softening of the $\nu_2$-mode at $H^{*}$ is essentially different. The lower panel of Fig.~\ref{spectra} illustrates the orientations of the spin plane and the polarization vectors before and after spin-flop in the presence of both $H$~and~$E$.

From Eqs.~(\ref{ptry})-(\ref{ptrz}) we obtain that $\tilde{\cal{P}}$ is expected to be (i) independent of the sign of the magnetic field $H,$ (ii) dependent on the sign of the electric polarization $P$, and (iii) proportional to the amplitude of the alternating electric field $\tilde{E}.$ These expectations were proven experimentally (see Figs.~\ref{36_2}-\ref{17_2}). The absence of response on electric field directed along $\boldsymbol{c}$-axis at $H<H^{*}$ is also in agreement with symmetry analysis. 

Field scans of $\cal{P}$ and $\tilde{\cal{P}}$ presented in Figs.~\ref{36_2},~\ref{41_8}, \ref{42c} were measured in arbitrary but the same units, that allows to determine the shift of the magnetic resonance field ${\Delta}H_{R}$ and, using Eqs.~(\ref{deltaY}),~(\ref{deltaZ}), to evaluate the electric polarization of the sample. The estimated value of spontaneous polarization along crystallographic axis $\boldsymbol{a}$ is equal to $P=18\pm 4$~$\mu$C/m$^2$. This value was obtained for electric field evaluated as $E=U/d$. Taking into account the nonuniformity of distribution of $E$ in the sample due to specific geometry of the samples (see Fig.~\ref{cell}b) the obtained value of $P$ should be considered as an estimation from below.
Extrapolated data of the temperature dependencies of the electric polarization obtained with pyrocurrent technique in Ref.~\cite{yasui_2008} give the value of $P_x\approx40$~$\mu$C/m$^2$ at $T=1.3$~K.

Nonzero responses of the transmitted power on the alternating electric field detected at ${\nu=17.2}$~GHz were observed in magnetic fields higher than $H_{sf}$ (see Fig.~\ref{17_2}). This observation shows that in the presence of electric field in the spin-flopped phase the vector $\boldsymbol{n}$ of the spin plane is deviated from the magnetic field direction (see lower panel of Fig.~\ref{spectra}) and one of two magnetic domains with different directions of vector $\boldsymbol{n}$ starts to be preferable. This result indicates that with application of an electric field it is possible to build single-domain magnetic state within spiral phase in LiCuVO$_4$ at magnetic fields before and after the spin-flop reorientation.

Finally, it should be mentioned that we did not observe the response on electric field in the more intriguing part of phase diagram of LiCuVO$_4$, where the chiral magnetic phase with zero magnetic moments of ions and ordered $\boldsymbol{n}$ is expected. Such phase was suggested in Ref.~\cite{ruff_2019} from the results of pyrocurrent measurements. Our evaluations have shown that sensitivity of our method was not enough to observe such response. 

\section{Conclusion}
The shift of the magnetic resonance spectra in presence of the electric field within the low-field multiferroic magnetic state of LiCuVO$_4$ has been observed and studied experimentally.
Symmetry analysis has been conducted in order to describe the static properties of the magnetic system. The low-frequency dynamics of LiCuVO$_4$ in magnetic and electric fields has been considered in the framework of hydrodynamic approach. Satisfactory agreement between the experimental results and the theoretical consideration has been obtained. 

It was shown that the magnetic structure of LiCuVO$_4$ can be efficiently controlled by both magnetic and electric fields, in the same time the magnetic structure of the sample can be checked by ESR technique discussed in this paper. These options can also be attractive for applications.  

\acknowledgments

We thank A.I.~Smirnov and H.-A.~Krug von Nidda for stimulating discussions.
Theoretical part of this work was supported by Russian Foundation for Basic Research No. 19-02-00194. Experimental part was supported by Russian Science Foundation Grant No. 17-12-01505.

\end{document}